\documentclass[12pt,english,british]{article}
\usepackage[T1]{fontenc}
\usepackage[koi8-r]{inputenc}
\usepackage{a4wide}
\usepackage{babel}
\usepackage{graphics}
\usepackage{setspace}
\makeatletter

\providecommand{\LyX}{L\kern-.1667em\lower.25em\hbox{Y}\kern-.125emX\@}
\let\SF@@footnote\footnote
\def\footnote{\ifx\protect\@typeset@protect
    \expandafter\SF@@footnote
  \else
    \expandafter\SF@gobble@opt
  \fi
}
\expandafter\def\csname SF@gobble@opt \endcsname{\@ifnextchar[
  \SF@gobble@twobracket
  \@gobble
}
\edef\SF@gobble@opt{\noexpand\protect
  \expandafter\noexpand\csname SF@gobble@opt \endcsname}
\def\SF@gobble@twobracket[#1]#2{}

 \newenvironment{lyxlist}[1]
   {\begin{list}{}
     {\settowidth{\labelwidth}{#1}
      \setlength{\leftmargin}{\labelwidth}
      \addtolength{\leftmargin}{\labelsep}
      }}
   {\end{list}}

\makeatother
\begin{document}

\title{Precision measurements of timing characteristics of the 8'' ETL9351
series photomultiplier}

\maketitle

\begin{center}
\author{\bf{
O.Ju.Smirnov\footnotemark[1],
P.Lombardi\footnotemark[2], 
G.Ranucci\footnotemark[2]}}
\end{center}
\footnotetext[1]{Corresponding author: Joint Institutr for Nuclear Research, 
141980 Dubna, Russia. E-mail: osmirnov@jinr.ru;smirnov@lngs.infn.it} 
\footnotetext[2]{Dipartimento di Fisica Universit\`a and I.N.F.N., Milano, Via Celoria, 16 I-20133 Milano, Italy}

\begin{abstract}
The results of the test measurements of the characteristics of 2200
PMT for the Borexino experiment provide the most complete information
for the evaluation of the ETL9351 timing characteristics with a high
precision. The unique timing characteristics of the apparatus used
and the large statistics accumulated during the tests of the PMTs
to be used in the future Borexino experiment, allow to resolve a fine
structure of the PMT timing response.

A method to obtain the probability density function of the single
photoelectron counting from the experimental data is proposed and
applied to derive the PMT average characteristics. For the first time,
an analytical model of the single photoelectron PMT time response
is proposed, describing all the features of the single photoelectron
time arrival. The origin of the small amplitude pulses is explained.
\end{abstract}

\newpage
\section{Introduction}

Photomultipliers with a large area spherical photocathode are being
widely used in many liquid scintillator and water cherenkov rare events
detectors. All recently developed solar neutrino detectors, such as
SNO \cite{SNO}, KamiokaNDE \cite{KamiokaNDE}, KamLAND \cite{KamLand},
Borexino \cite{Borex} and its Counting Test Facility \cite{CTF},
are based on the scintillation photons counting techniques. The interaction
point in these detectors is reconstructed using timing information
from a large number of PMTs. Depending on the precision of reconstruction
and the total number of PMTs the precision of single photoelectron
detection at the level of 1 ns is demanded. The Monte Carlo simulation
of the Borexino detector showed that the mean number of photoelectrons
(p.e.) registered by one PMT in a scintillation event will be in the
region \( 0.02-2.0 \) for an event with energy of 250-800 keV. Hence
the PMTs should demonstrate a good single electron performance. After
preliminary tests, the ETL 9351 with a large area photocathode (8'')
has been chosen \cite{ETL}. The PMT of this model has 12 dynodes
with a total gain of \( k=10^{7} \). The transit time spread of the
single p.e. response is \( 1-1.5 \) ns. The PMT has a good energy
resolution characterized by the manufacturer by the peak-to-valley
ratio. The manufacturer (Electron Tubes Limited, ETL) guarantees a
peak-to-valley ratio of 1.5. The results of the preliminary tests
with 50 PMTs have been reported in \cite{PMTcharact}.

\section{PMT test facility at LNGS}

In the Borexino programme the special PMT test facility was prepared
at LNGS. The test facility is placed in two adjacent rooms. In one
room the electronics is mounted, and the other is a dark room with
4 tables designed to hold up to 64 PMTs. The dark room is equipped
with an Earth's magnetic field compensation system using rectangular
coils with an electric current (\cite{MField}). The non-uniformity
of the compensated field in the plane of the tables is no more than
\( 10\% \). The tables are separated from each other by black shrouds,
which screen any light reflected from the PMTs photocathode.

\begin{figure*}[!thhhh]
{\centering \resizebox*{1\textwidth}{0.5\textheight}{\includegraphics{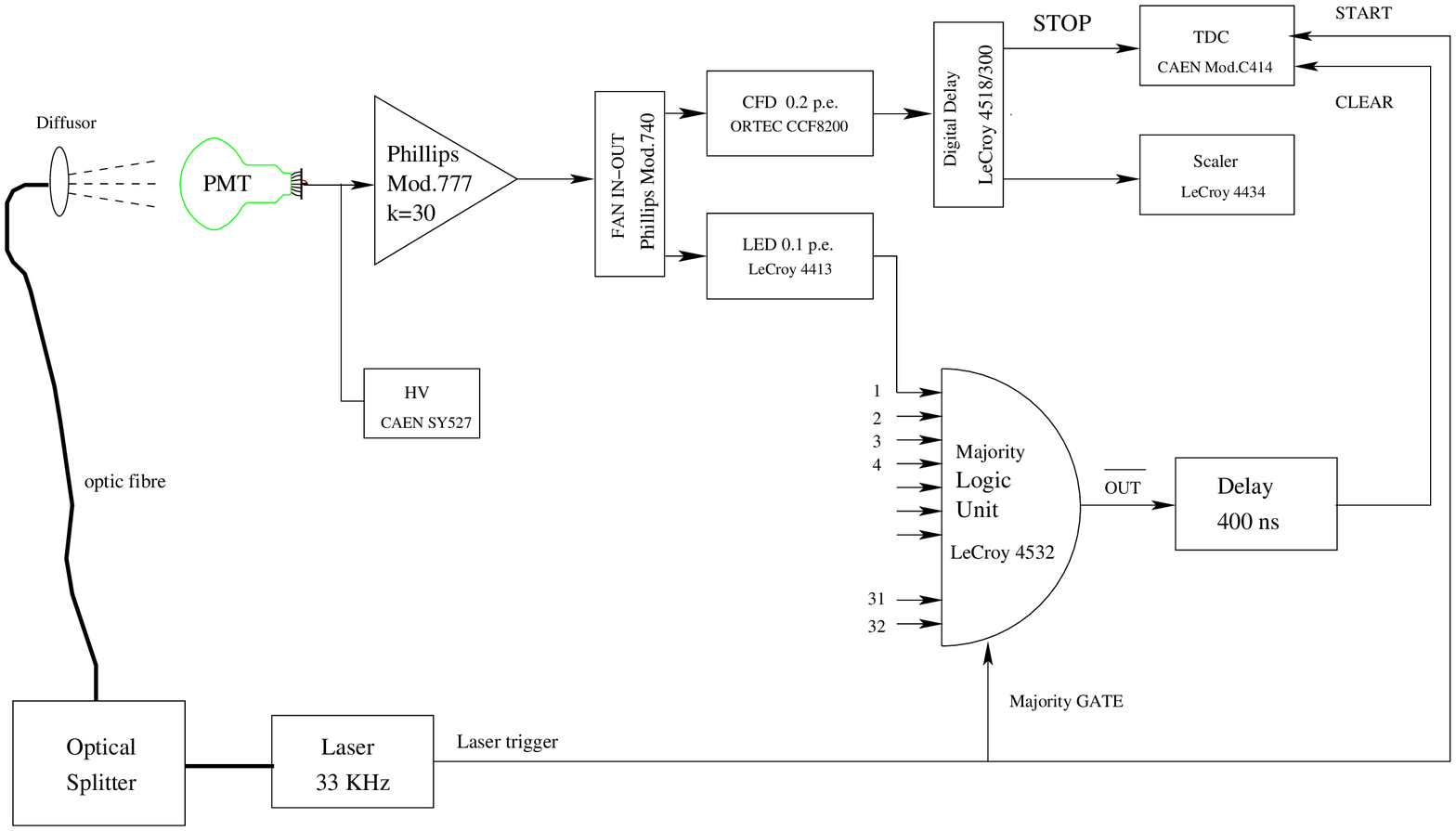}} \par}

\caption{\label{Figure:Electronics}The simplified scheme of the one channel
of the electronics.}
\end{figure*}

The simplified scheme of one channel of electronics (out of the total
32) of the test facility is presented in Fig.\ref{Figure:Electronics}.
The system uses the modular CAMAC standard electronics and is connected
to a personal computer by the CAEN C111 interface. The PMT characteristics
are normally measured during a 5 hour run.

The PMTs are illuminated by low intensity light pulses from a laser.
A picosecond Hamamatsu pulse laser was used in the tests. The model
used has a peak power of \( 0.39\: mW \), the pulse width is \( 27.3\: ps \),
and the laser wavelength is \( 415\: nm \), which is close to the
maximum sensitivity of the ETL 9351 photocathode. The light pulse
from the laser is delivered by 6 meter long optical fibers into the
dark-room. Each of the 4 fibers is supplied with a diffuser in order
to provide a more uniform illumination of the tables.

The TDC, CAEN mod.C414, {}``start'' signals are generated using
the laser internal trigger\c{ } which has negligible time jitter
(\( <100\: ps \)) with respect to the light pulse. The {}``stop''
signal for the TDC is formed by the constant fraction discriminator
(CFD, ORTEC mod.CCF8200) with the threshold set at the 0.2 p.e. level.
The full scale of TDC was set to 200 ns with 2048 channels resolution.
Because of the memory restrictions of the software, only the part
of the full range was used, namely 100 ns in the region {[}-30 ns;+70
ns{]} around the main peak in the PMT transit time.

The 32-input majority logic unit, LeCroy mod.4532, is able to memorize
the pattern of the hit channels. This information significantly increases
the data processing rate. The reading of the electronics is activated
when the majority LAM signal is on (a LAM is produced if one of the
signals on the input is inside the external GATE on the majority logic
unit). Otherwise, a hardware clear is forced using the majority \( \overline{OUT} \)
signal. Every pulse of the laser is followed by an internal trigger.
The trigger is used as the majority external gate.

A high precision calibration of each electronics channel had been
performed before the measurements. Here calibration means the precise
knowledge of the response to a signal corresponding to 1\( \:  \)p.e.%
\footnote{multiplied by a factor of \( 10^{7} \) by the electron multiplier
and giving \( 1.6\:  \)pC charge
} on the system input. The PMT in this measurement was substituted
by a precision charge generator LeCroy mod. 1976.

The gain of each PMT electron multiplier was set to a value of \( 2\cdot 10^{7} \),
before the tests with a help of automated gain adjustment system,
described in \cite{HV}.

\section{\label{Section:Testing}Results of the 2000 PMT testing}

The main timing characteristics defined during the acceptance tests
were:

\begin{lyxlist}{00.00.0000}
\item [\textbf{\( t_{0}\: and\: \sigma _{t} \)}]the position and the rms
of the gaussian fitting the main peak in the transit time distribution.
The fit had a an additional constant corresponding to the dark noise
level;
\item [\( rms \)]is estimated for all transit time histogram (up to 90
ns);
\item [\( p_{late} \)]late pulsing in percent, estimated as the ratio of
the events in the \( [t_{0}+3\cdot \sigma _{t};100] \) ns range to
the total number of the events;
\item [\( p_{prep} \)]prepulsing in percent, estimated as the ratio of
the events in the \( [0;t_{0}-3\cdot \sigma _{t}] \) ns range to
the total number of the events.
\end{lyxlist}
The results of measurements showed no essential problems with the
transit time spread of the PMTs, with PMTs rejected mainly or because
of the high dark rate or because of the bad amplitude response of
the PMT. As a rule a PMT with a good single photoelectron charge response
has also a good timing response. 

The results of the measurements are presented in Table \ref{Table:Tests}.
For future use we put also in this table the parameter \( p_{U} \),
which is the fraction of the underamplified signals in the amplitude
spectrum of the single photoelectron response. The model used to extract
the value for \( p_{U} \) from the single p.e. charge response is
described in \cite{Filters}. The underamplified signals can be described
well with an exponential with a negative slope \( A=-0.17 \) p.e.,
the value of \( A \) is presented in Table \ref{Table:Tests} too.
The measurements with a threshold (Th) set to 0.16 p.e. cuts 61\%
of the underamplified signals, leaving 6.4\% of the total 16.5\%. 

The distribution of the \( t_{0} \), \( \sigma _{t} \) and \( rms \)
parameters is a normal distribution with a sigma coinciding with the
rms of distribution. The distributions for \( p_{late} \) and \( p_{prep} \)
have longer non-gaussian tails. All these results have been written
in a database, which can be used for the detector modeling.

\begin{table*}[!thhhh]

\caption{\label{Table:Tests}The results of test.}

{\centering \begin{tabular}{|c|c|c|c|c|c|c|c|c|}
\hline 
\selectlanguage{english}
parameter
\selectlanguage{british}&
\selectlanguage{english}
\( t_{0} \)
\selectlanguage{british}&
\selectlanguage{english}
\( \sigma _{t} \)
\selectlanguage{british}&
\selectlanguage{english}
\( rms \)
\selectlanguage{british}&
\selectlanguage{english}
\( p_{late} \)
\selectlanguage{british}&
\selectlanguage{english}
\( p_{prep} \)
\selectlanguage{british}&
\selectlanguage{english}
\( p_{U} \)
\selectlanguage{british}&
\selectlanguage{english}
A
\selectlanguage{british}&
\selectlanguage{english}
Th
\selectlanguage{british}\\
&
ns&
ns&
ns&
\%&
\%&
&
p.e.&
p.e.\\
\hline 
\selectlanguage{english}
mean
\selectlanguage{british}&
\selectlanguage{english}
32.59
\selectlanguage{british}&
\selectlanguage{english}
1.18
\selectlanguage{british}&
\selectlanguage{english}
8.14
\selectlanguage{british}&
\selectlanguage{english}
7.27
\selectlanguage{british}&
\selectlanguage{english}
0.69
\selectlanguage{british}&
\selectlanguage{english}
0.165
\selectlanguage{british}&
\selectlanguage{english}
-0.17
\selectlanguage{british}&
\selectlanguage{english}
0.16
\selectlanguage{british}\\
\hline 
\selectlanguage{english}
rms
\selectlanguage{british}&
\selectlanguage{english}
3.85
\selectlanguage{british}&
\selectlanguage{english}
0.11
\selectlanguage{british}&
\selectlanguage{english}
0.54
\selectlanguage{british}&
\selectlanguage{english}
1.0
\selectlanguage{british}&
\selectlanguage{english}
0.30
\selectlanguage{british}&
\selectlanguage{english}
0.05
\selectlanguage{british}&
\selectlanguage{english}
0.068
\selectlanguage{british}&
\selectlanguage{english}
0.04
\selectlanguage{british}\\
\hline
\end{tabular}\par}
\end{table*}

\begin{figure*}[!thhhh]
{\centering \resizebox*{1\textwidth}{0.5\textheight}{\includegraphics{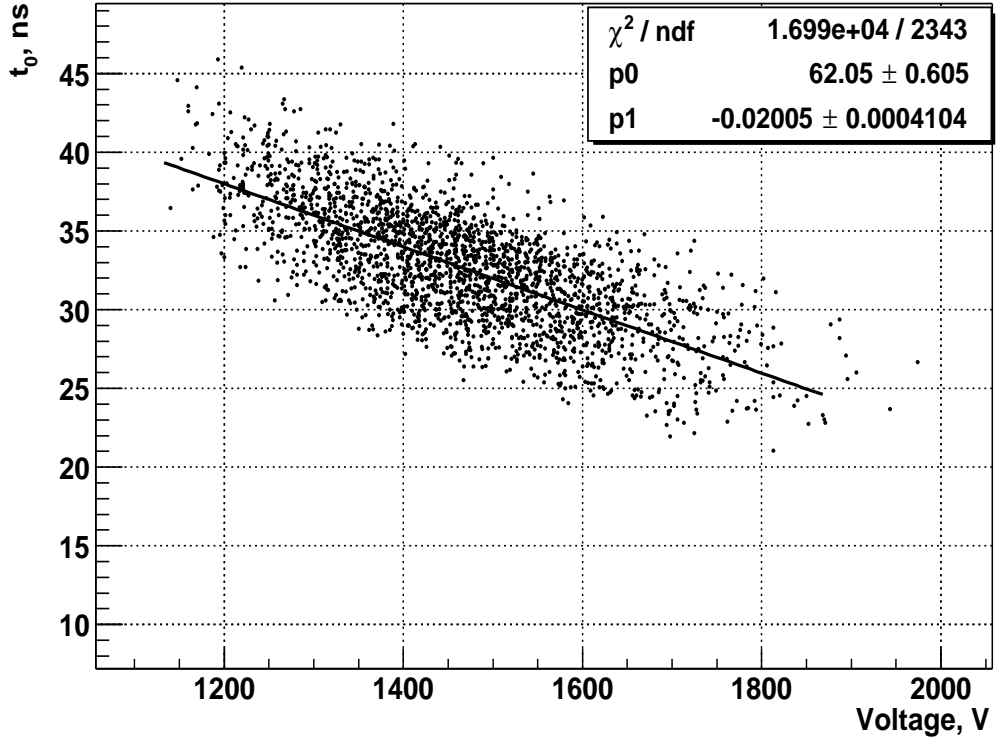}} \par}

\caption{\label{Figure:t0vsHV}The dependence of the transit time peak on
the applied voltage.}
\end{figure*}

An important characteristic of the PMT is the dependence of the peak
of the transit time \( t_{0} \) on the applied voltage, see Fig.\ref{Figure:t0vsHV}.
The shift of the \( t_{0} \) position due to a change in the voltage
applied is \( 0.02 \) ns/V. This value can be used in order to equalize
the time arrival of photoelectrons after each adjustment of the high
voltage.

\section{The averaged transit time shape}

The straightforward use of the database, with a set of parameters
for each PMT, for the detector's modeling will unnecessary slow down
the calculations, because of the huge number of PMTs (2200) used in
the experiment. The calculation speed can be improved using average
characteristics of the PMTs instead of the individual ones. Because
every PMT operates at its own voltage, and the lighting conditions
depends on the position on the test tables, the procedure of averaging
should be preceded by the equalizing the difference in the conditions.
Fortunately, the statistics of the first photoelectron counting provides
this possibility.

\subsection{Statistics of the first photoelectron time arrival }

With any experimental conditions, the PMT does not register single
photoelectron, and so the single photoelectron response should be
extracted from a PMT response \cite{Ranucci1,Ranucci2}. The basic
assumption in the following considerations is a Poisson distribution
for the amount of the registered photoelectrons.

If the probability density function (p.d.f.) of a single p.e. registering
at time \( t \) is \( \rho _{1}(t) \), then the probability to observe
the first p.e. out of precisely \( n \) photoelectrons at time \( t \)
is

\begin{equation}
\label{Formula:RhoN}
\rho _{n}(t)=n\cdot \rho _{1}(t)\cdot (1-F_{1}(t))^{n-1}
\end{equation}

where \( F_{1}(t)=\int _{-\infty }^{t}\rho _{1}(t)dt \) is probability
to observe a single p.e. before time \( t \).

The equivalence of the photoelectrons gives the factor \( n \), and
\( (1-F(t))^{n-1} \) is the probability of not observing any of the
remaining \( (n-1) \) p.e. before time \( t \).

If the number of photoelectrons is not fixed, but distributed in accordance
to a Poisson law with a mean \( \mu  \) p.e. per pulse, then the
probability of the arrival of the first signal at the PMT can be calculated
by averaging (\ref{Formula:RhoN}) over the Poisson distribution \( P(n)=\frac{\mu ^{n}}{n!}e^{-\mu } \):

\textbf{\begin{equation}
\label{RhoT}
\rho (\mu ,t)=\mu \rho _{1}e^{-\mu F_{1}(t)}.
\end{equation}
}The p.d.f. in (\ref{RhoT}) is normalized by the total probability
of the presence of a non-zero signal for a Poissonian distribution
\( P(n>0)=1-e^{-\mu } \) :

\[
\int _{-\infty }^{+\infty }\rho (\mu ,t)dt=1-e^{-\mu },\]

which can be easily checked out noting that \( \rho _{1}(t)dt=dF(t) \). 

The full probability to register a signal in the interval \( [-\infty ,t] \)
is then:

\begin{equation}
\label{F(T)}
F(\mu ,t)\equiv \int _{-\infty }^{t}\rho (\mu ,t')dt'=1-e^{-\mu F_{1}(t)}.
\end{equation}

From (\ref{RhoT}) and (\ref{F(T)}) follows a simple relation:

\begin{equation}
\label{Rho1_exp}
\rho _{1}(t)=\frac{1}{\mu }\frac{\rho (\mu ,t)}{1-F(\mu ,t)}.
\end{equation}

which allows to calculate the \( \rho _{1}(t) \) function using experimental
data.

If experimental data are presented in the form of a histogram \( N_{Exp}(i) \),
then the probability density function of the single photoelectron
can be calculated in accordance with (\ref{Rho1_exp}):

\begin{equation}
\label{Formula:Hist}
N_{1}(i)=\frac{1}{\mu }\frac{N_{Exp}(i)}{1-s\; (i)},
\end{equation}

where \begin{equation}
\label{Formula:RunningSum}
s\: (i)\equiv \frac{1}{N_{Triggers}}\sum _{k=1}^{k=i}N_{Exp}(k)
\end{equation}
is the running sum of the histograms of the experimental data \( N_{Exp}(i) \)
normalized by the number of the system starts \( N_{Triggers} \).
Naturally, when \( N_{Triggers} \) is large enough, one can expect
\( s\: (\infty )=1-e^{-\mu } \).

For completeness, let us give the equation for the estimation of the
errors in \( N_{1}(i) \):

\[
\sigma _{1}(i)=\frac{1}{\mu }\frac{\sigma (i)}{1-s\: (i)}.\]

The equation (\ref{Rho1_exp}) is especially useful when calculating
the function \( \rho _{1}(t) \) from the experimental data with \( \mu \simeq 1 \),
where the shape of \( \rho (t) \) is significantly different from
the shape of \( \rho _{1}(t) \), and the approximation \( \rho (t)\simeq \mu \rho (t) \)
can not be considered satisfactory. In our measurements the \( \mu \simeq 0.05 \)
and the correction applied is of the same order at the distribution
tail (at small \( \mu  \) the approximation \( F(t_{Max})\simeq \mu  \)
is valid).

\subsection{Correcting for the random coincidence with a dark noise}

The laser system has been tuned to provide mean counting rate of the
PMTs \( \mu \simeq 0.05 \) p.e.; this condition ensures a mostly
single p.e. regime for the PMT (with the relative probabilty of the
signal originating from 2 p.e. and more \( r=\frac{P(n>1)}{P(n>0)}=\frac{1-e^{-\mu }-\mu e^{-\mu }}{1-e^{-\mu }}\simeq  \)\( \frac{\mu }{2} \),
i.e. 2.5 \%). With such a small amount of light in a pulse, the PMT
response could be affected by the dark noise of the PMT, which is
of the order of some kHz. The probability of the random coincidences
due to the dark noise \( f_{dark} \) in the time window \( \tau  \)
can be expressed by

\[
f_{rndm}=f_{dark}\cdot \tau ,\]

and the total amount of dark noise counts in each bin of the histogram
(of the total \( N_{bins} \)) is

\[
N_{dark}(i)=N_{Triggers}\frac{f_{rndm}}{N_{bins}}.\]

In order to take correctly into account the random noise in the system
one should substitute \( N_{exp}(i) \) by \( N_{exp}(i) \)-\( N_{dark}(i) \)
in equations (\ref{Formula:Hist}) and (\ref{Formula:RunningSum}).

The dark noise in the system is measured independently with high precision
using scalers.

\subsection{The procedure used to obtain the mean characteristics of the PMT}

\begin{figure*}[!thhhh]
{\centering \resizebox*{1\textwidth}{0.5\textheight}{\includegraphics{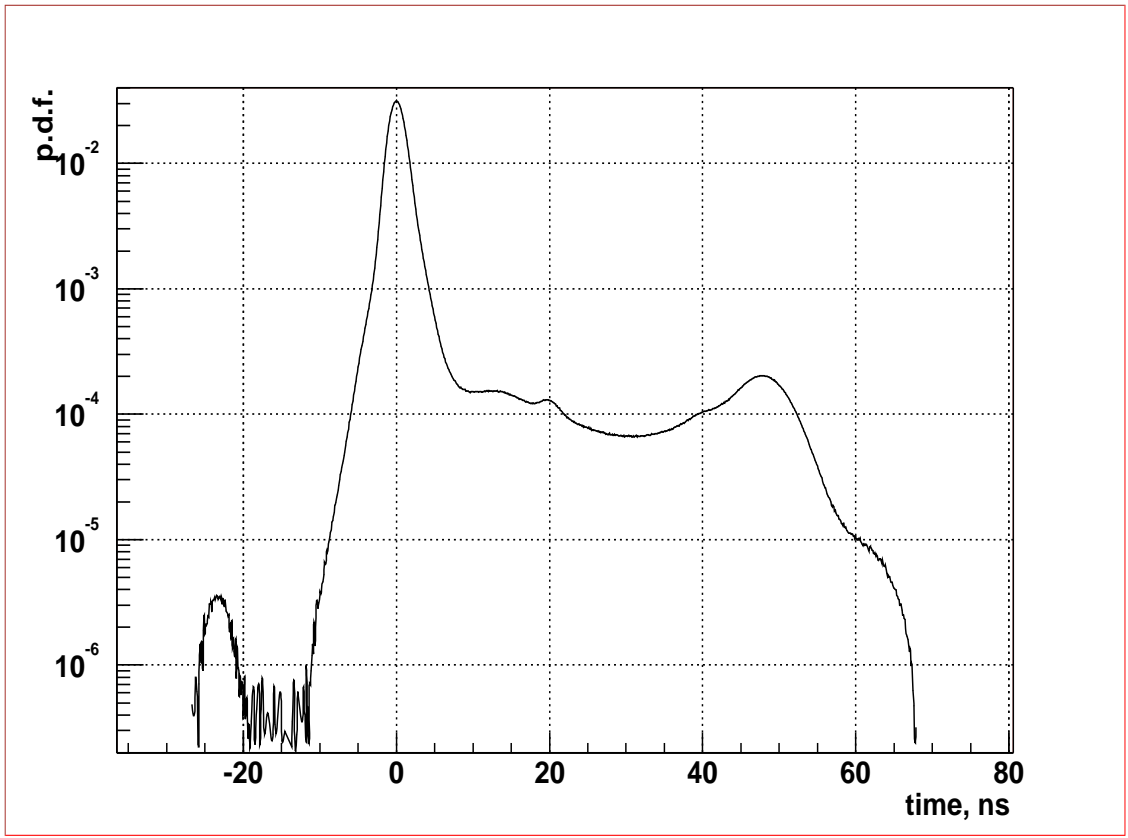}} \par}

\caption{\label{Figure:TDC}The averaged timing characteristics of the ETL9351
PMT. }
\end{figure*}

\begin{enumerate}
\item Using the measured value of the dark rate the contribution \( N_{dark} \)
of the random coincidences at one bin was calculated.
\item Using equations (\ref{Formula:Hist}) and (\ref{Formula:RunningSum})
with \( N_{exp}(i) \) substituted by \( N_{exp}(i) \)-\( N_{dark}(i) \)
the \( N_{1}(i) \) function was calculated and normalized: \begin{equation}
\label{Formula:n1}
n_{1}=\frac{N_{1}(i)}{\sum _{i=1}^{N_{bins}}N_{1}(i)}.
\end{equation}
As follows from (\ref{Formula:n1}), the knowledge of the mean number
of the registered photoelectrons is not necessary for the calculation
of the normalized probability.
\item The peak in the distribution \( n_{1} \) is found and the histogram
is shifted in order to put its maximum at the position corresponding
to \( t=0 \). 
\item All the histograms are summed together and normalized to 1 once more.
The obtained histogram contains the mean characteristics of the sample
of the PMTs used with a peak (not mean time of the arrival) at the
position \( t=0 \).
\end{enumerate}
The resulting histogram is presented in Fig.\ref{Figure:TDC}. This
is the PMT transit time p.d.f. averaged over a 2000 PMT sample.

\section{The structure of the photomultiplier transit time}

The following features of the transit time curve can be clearly seen
in Fig.\ref{Figure:TDC}: 1)almost gaussian peak at the position \( t=0 \)
ns; 2)a very weak peak at \( t=-24 \) ns; 3)a weak peak at \( t=48 \)
ns; 4)the continuous distribution of the signals arriving between
the main peak and the peak at \( t=48 \) ns; 5)another very weak
peak at \( t=20 \) ns. All these features, together with others appearing
with closer investigation of the curve, will be explained in the current
section.

\subsection{Main peak}

\begin{figure*}[!thhhh]
{\centering \resizebox*{1\textwidth}{0.5\textheight}{\includegraphics{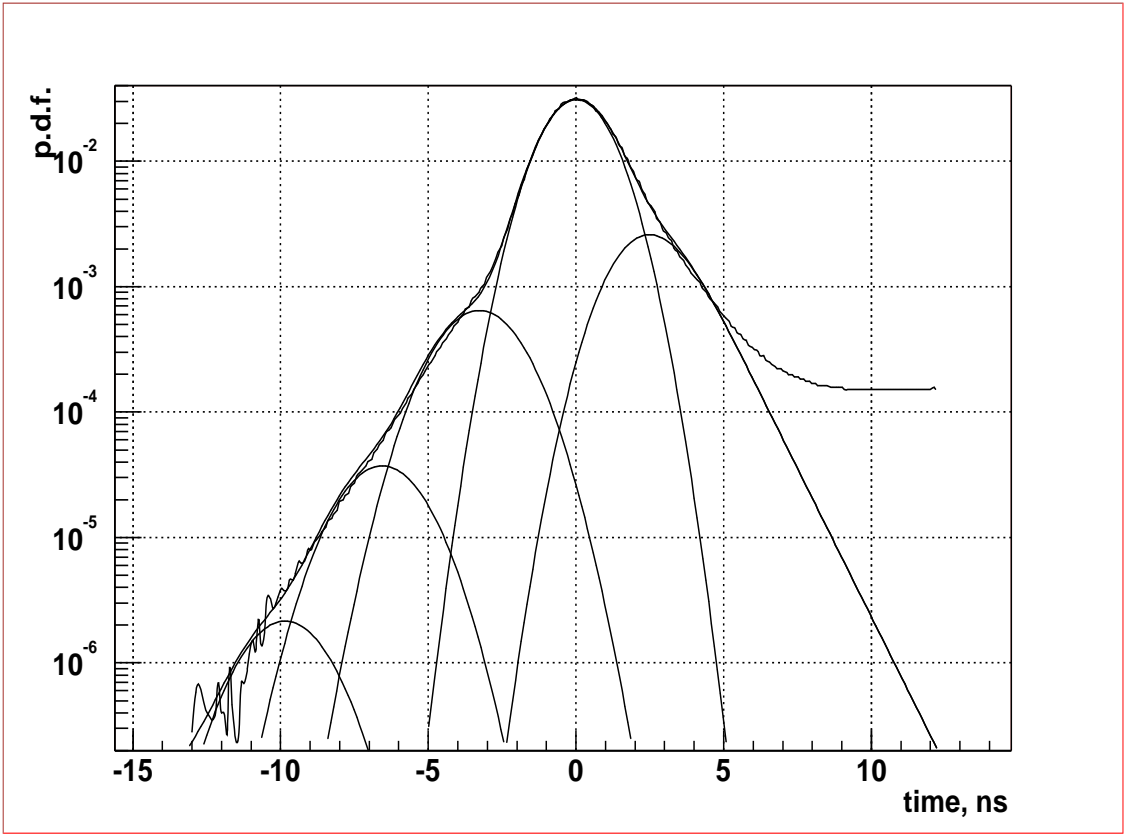}} \par}

\caption{\label{Fig:MainPeak}Main peak region}
\end{figure*}

The fine structure of the main peak can be seen in Fig.\ref{Fig:MainPeak}.
The main peak has almost gaussian shape, but at the regions \( t>2 \)
ns and \( t<-2 \) ns the deviation from the gaussian distribution
is significant. Early pulses can arrive due to the elastic scattering
of the photoelectron on the first dynode without multiplication, in
such a way arriving at the next stage of the electron multiplier earlier
than secondaries do. The same process can occur on the second dynode,
third etc. The energy of the electron arriving at the first dynode
is defined by the potential difference between the photocathode and
the first dynode \( U_{D1} \). The \( U_{D1} \) voltage is constant
in the divider scheme used, which is provided by three Zenner diodes
of 200 V each. The energy gained by the elastically scattered electron
at the second stage is small in comparison to the initial 600 \( eV \),
thus the velocity of the electron in the sequence of the elastic scatterings
can be considered constant in first approximation, as well as the
transit time between the first dynodes. The arriving of the early
signals can be modeled by a set of the equidistant gaussians with
the same spread and geometrically decreasing strength:

\begin{equation}
\label{Formula:EarlyPulsesSum}
f_{e}(t)=\frac{1}{p+p^{2}+..+p^{N}}\sum _{n=1}^{N}\frac{p^{n}}{\sqrt{2\pi }\sigma _{e}}e^{-\frac{1}{2}(\frac{t+n\cdot \delta _{t}}{\sigma _{e}})^{2}}.
\end{equation}

The pulses just after the main peak are due to the inelastic scattering
on the first dynode. 

Let us introduce a set of variables, describing the relative probabilities
of the considered processes: \( p_{g} \) is the relative probability
of the signals under the gaussian part of the peak; the early pulses
arrive with a relative probability \( p_{e} \) and the late pulses
have a relative probability \( p_{l} \). 

We will describe the late pulses using a function:

\begin{equation}
\label{Formula:ExpGauss}
f_{l}(t)=\frac{\sigma _{l}^{2}-(t-t_{l})\cdot \tau }{2\tau ^{2}}(1+erf(\frac{(t-t_{l})\cdot \tau -\sigma _{l}^{2}}{\sqrt{2}\cdot \tau \cdot \sigma _{l}}))
\end{equation}

which is a convolution of an exponential with a slope \( \frac{1}{\tau } \),
and a gaussian with sigma \( \sigma _{l} \).

The main peak fits well with the following function:

\begin{equation}
\label{Formula:MainPeak}
M(t)=p_{g}\cdot f_{g}(t-t_{0})+p_{e}\cdot f_{e}(t-t_{0})+p_{l}\cdot f_{l}(t-t_{0})
\end{equation}

\begin{figure*}[!thhhh]
{\centering \resizebox*{1\textwidth}{0.5\textheight}{\includegraphics{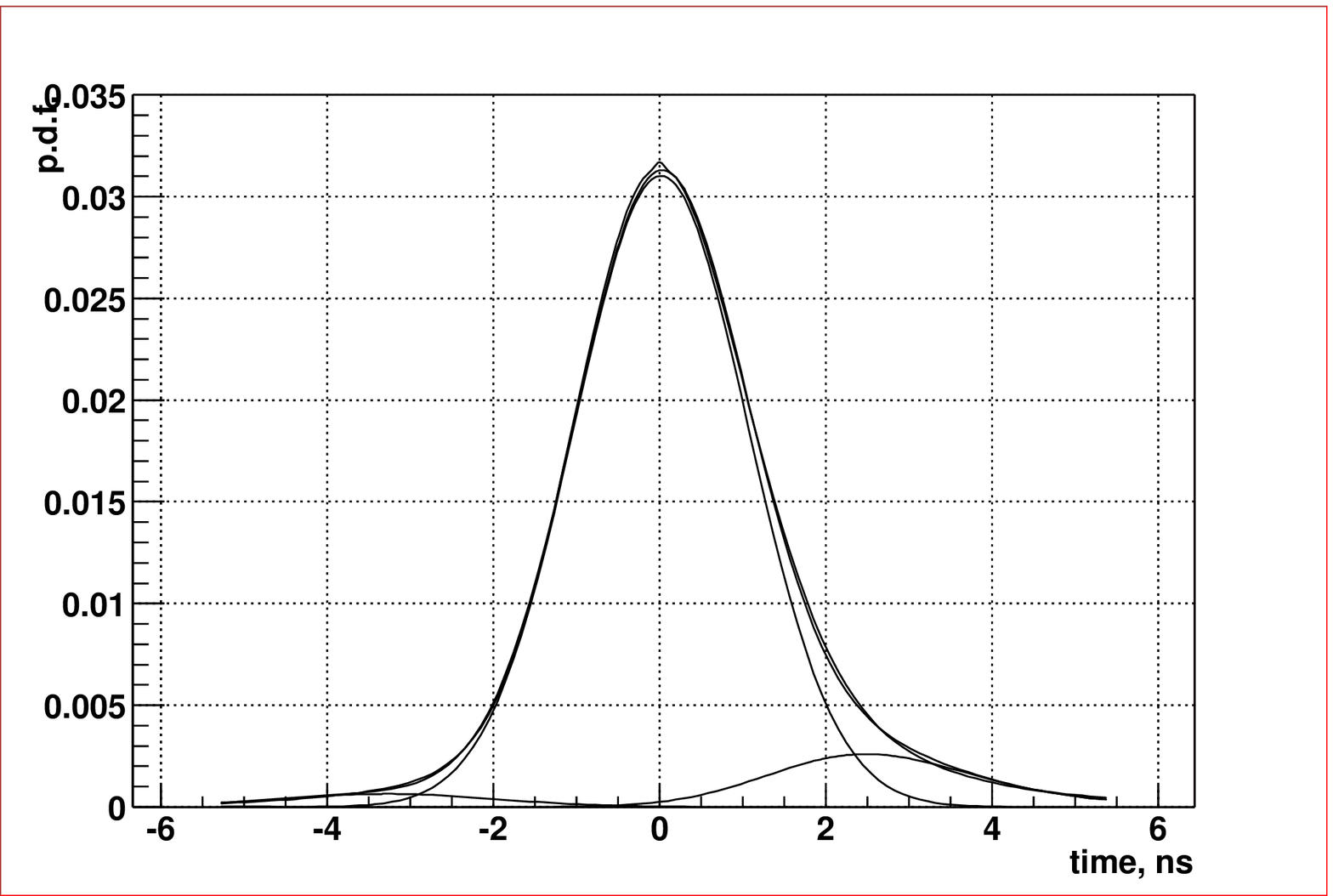}} \par}

\caption{\label{Figure:MainPeakLin}Main peak region on the linear scale.}
\end{figure*}

The results of the fit can be seen in Fig.\ref{Fig:MainPeak} and
Fig.\ref{Figure:MainPeakLin}. The parameters of the best fit are
presented in Table\ref{Table:BestFit}. The model of the main peak
describes \( 94 \) \% of all pulses.

\begin{table*}[!thhhh]

\caption{\label{Table:BestFit}The parameters of the best fit.}

{\centering \begin{tabular}{|c|c|c|c|c|c|c|c|c|c|}
\hline 
\( p_{g} \)&
\( p_{e} \)&
\( p_{l} \)&
\( t_{0} \)&
\( \sigma  \)&
\( p \)&
\( \delta _{e} \)&
\( \sigma _{e} \)&
\( \tau  \)&
\( t_{l} \)\\
\hline 
0.83&
0.023&
0.085&
0.02&
1.04&
0.06&
3.3&
1.29&
0.92&
1.80\\
\hline
\end{tabular}\par}
\end{table*}

The time of flight between the dynodes \( \delta _{t}=2.76 \) ns
corresponds to the difference in the time of flight of the fast electron
between the dynodes and the drift time of the secondary electrons.

The amplitude of the early pulses was not measured in our tests, but
some can be deduced from the model of the early pulsing. The coefficient
of the multiplication at energies \( E_{e}>600 \) eV is almost independent
on energy. Thus, a photoelectron missing the first stage of multiplication
(with mean gain \( g_{1} \)) and multiplied at the second stage with
a gain \( g_{1} \) instead of \( g_{2} \) will produce at the anode
a signal with a mean amplitude reduced by the factor \( f_{1}=\frac{1}{g_{2}} \),
which is normally in the range \( 0.2-0.3 \). The reduction factor
\( f_{2} \) for a photoelectron missing two first stages of multiplication
will be \( f_{2}=\frac{1}{g_{2}\cdot g_{3}} \), etc. As one can see
from Table.\ref{Table:BestFit} the geometrical progression factor
is much lower, \( p=0.06 \) because of the threshold effect. In order
to estimate the part of the signals over the threshold we note that
the underamplified signals have an exponential distribution. The decrease
in amplitude of the signal corresponds to the increase of the slope
of the exponential distribution. If the threshold is fixed, then the
part of the signals over the threshold is \( p=(e^{-\frac{q_{th}}{A}})^{g_{2}} \)
for the signals missing the first stage of amplification. \( A=0.17 \),
the mean threshold is \( q_{th}=0.16 \), and \( g_{2}\simeq 3 \),
hence \( p\simeq e^{-3}\simeq 0.05 \), in agreement with the fit
value. The same rule is valid for the electrons missing two and more
stages of amplification.

\subsection{Prepulses}

\begin{figure*}[!thhhh]
{\centering \resizebox*{1\textwidth}{0.5\textheight}{\includegraphics{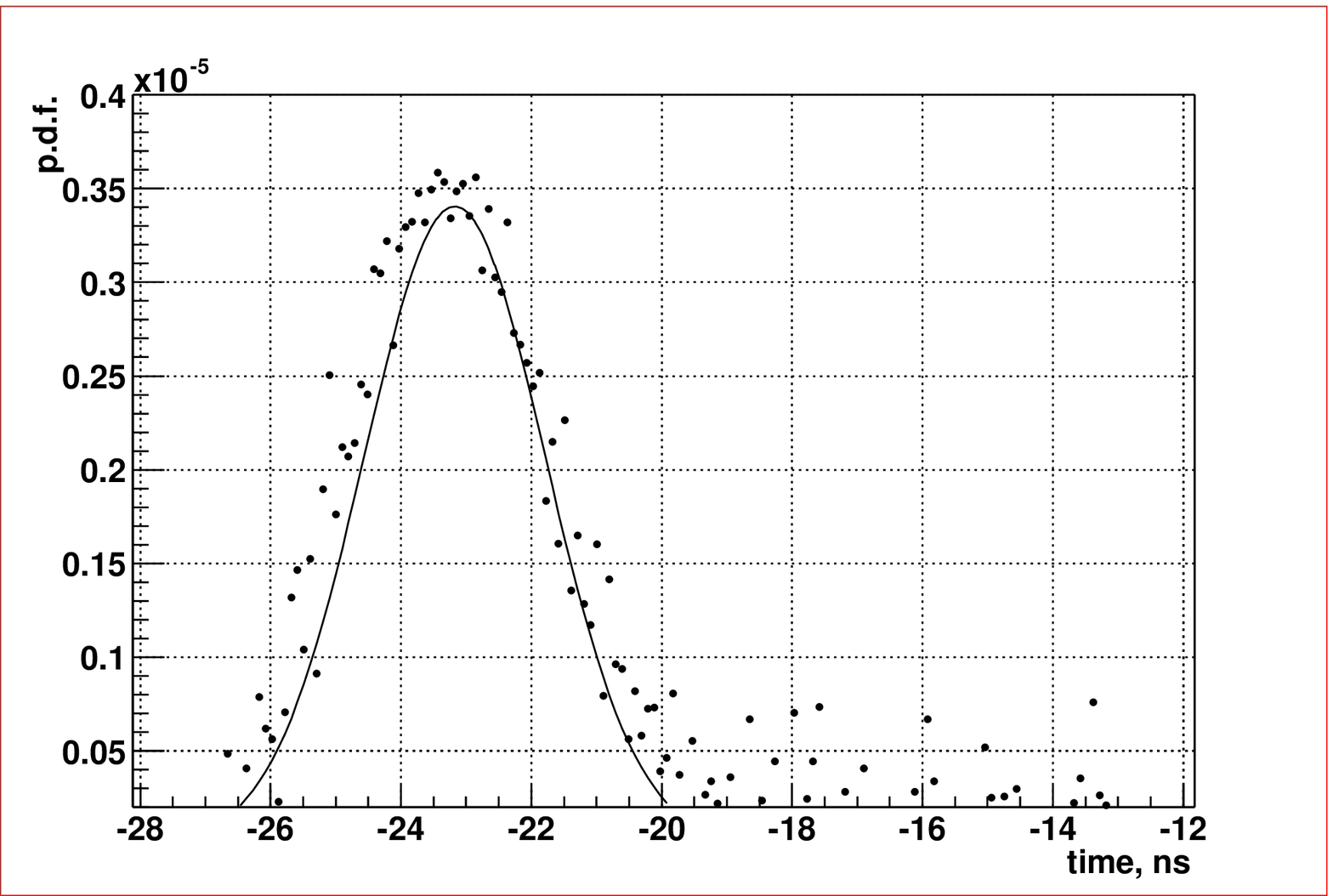}} \par}

\caption{\label{Figure:Prepulses}Prepulses}
\end{figure*}

In the transit time histogram can be clearly seen a small peak at
about -24ns. These are so called prepulses, corresponding to the direct
photoproduction of the electron on the first dynode. The amplitude
of these pulses is factor \( g_{1} \)(amplification of the first
dynode) less then the amplitude of the main peak. Because a typical
value is \( g_{1}\simeq 10 \), these pulses are strongly suppressed
by the CFD threshold set at the 0.2 p.e. level. The shape of the peak
is well approximated by a gaussian (see Fig.\ref{Figure:Prepulses})
with a parameters given in Table \ref{Table:Prepulses}.

\begin{table*}[!thhhh]

\caption{\label{Table:Prepulses}The parameters of the prepulses peak.}

{\centering \begin{tabular}{|c|c|c|}
\hline 
\selectlanguage{english}
\centering \( p_{pp} \)
\selectlanguage{british}&
\selectlanguage{english}
\( t_{pp} \)
\selectlanguage{british}&
\selectlanguage{english}
\( \sigma _{pp} \)
\selectlanguage{british}\\
\hline 
\selectlanguage{english}
\( 1.22\cdot 10^{-4} \)
\selectlanguage{british}&
\selectlanguage{english}
-23.18
\selectlanguage{british}&
\selectlanguage{english}
1.39
\selectlanguage{british}\\
\hline
\end{tabular}\par}
\end{table*}

The difference \( dt=23.2 \) ns between the position of the main
peak \( t_{0} \), and the position of the prepulses peak \( t_{pp} \)
corresponds to the drift time of the electron from the photocathode
to the first dynode \( t_{d} \) with the time of flight of photon
to the first dynode \( t_{tof} \) subtracted: \( dt=t_{0}-t_{pp}=t_{d}-t_{tof} \)
The time of flight can be calculated from the known distance between
the photocathode and the first dynode, which is 123 mm (radius of
the spherical photocathode is 110 mm, the focusing grid is situated
at the center of the sphere, the distance between the focusing grid
and the first dynode is 13 mm). Hence the time of flight of photon
inside the PMT is \( tof=0.41 \) ns, and the drift time \( t_{d}=dt+tof=23.61 \)
ns. The drift time is the same for all the PMTs tested, because the
potentials difference between the photocathode and the first dynode
is stabilized.

In setups with a large number of PMT in use, the presensce of prepulses
is a potential source of the early triggers in the system.

\subsection{Late pulses}

\begin{figure*}[!thhhh]
{\centering \resizebox*{1\textwidth}{0.5\textheight}{\includegraphics{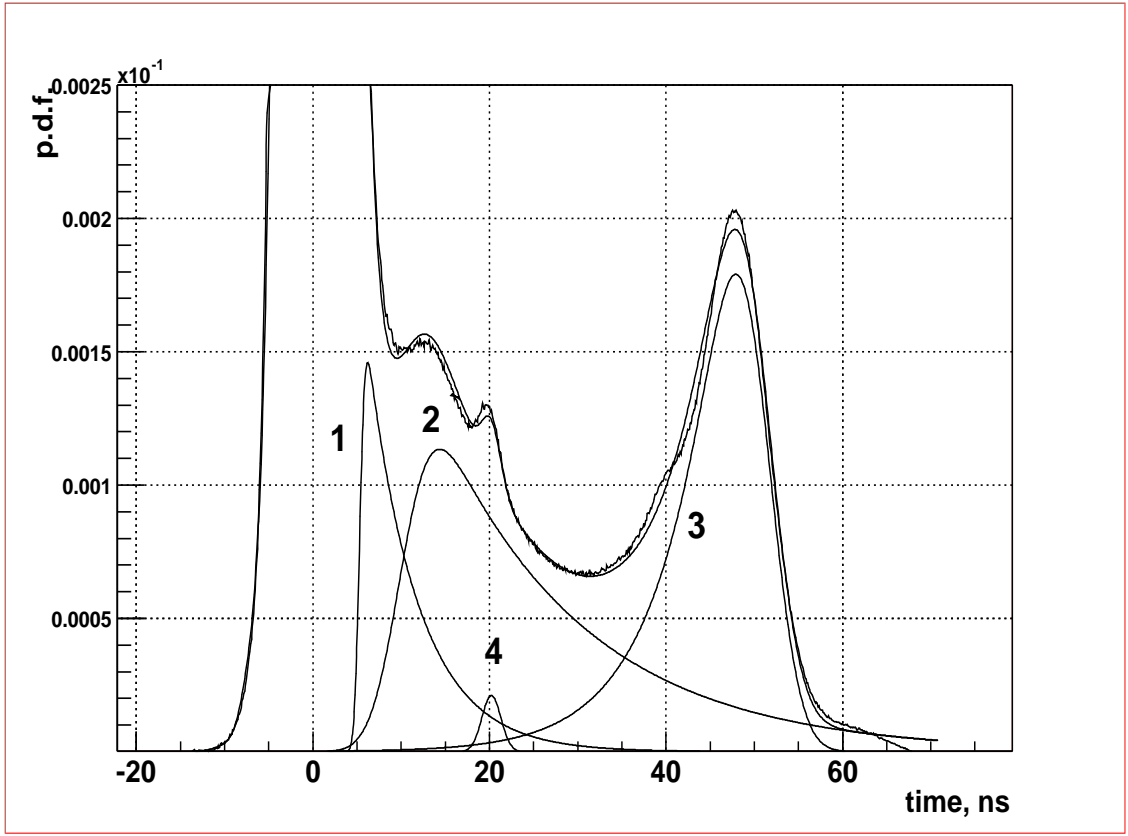}} \par}

\caption{\label{Figure:LatePulses}Late pulses. Lines 1 and 2 corresponds
to inelastically scattered phtoelectrons, curve 3 describes elastically
scattered photoelectrons. Gaussian shape 4 describes light reflections
in the light splitting system.}
\end{figure*}

Pulses arriving after the main pulse, in the time range up to 100
ns, are called late pulses. The structure of the late pulses spectrum
can be seen in Fig.\ref{Figure:LatePulses}. We are not considering
here pulses arriving in the microseconds interval, which are caused
by the ion-feedback. In literature these pulses are called afterpulses. 

The shape of the late pulses is modeled well with a sum of 3 functions
of the type given in equation (\ref{Formula:ExpGauss}).

The parameters of the best fit are presented in Table \ref{Table:LatePulsesFit}.
It should be noted that the function \( M(t) \) describing the main
peak has been fixed during the fit of the late pulses shape.

\begin{table*}[!thhhh]

\caption{\label{Table:LatePulsesFit}Late pulses fit parameters}

{\centering \begin{tabular}{|c|c|c|c|c|}
\hline 
\multicolumn{1}{|c|}{component}&
\selectlanguage{english}
\( p_{1} \)
\selectlanguage{british}&
\selectlanguage{english}
\( t_{1} \)
\selectlanguage{british}&
\selectlanguage{english}
\( \sigma _{1} \)
\selectlanguage{british}&
\selectlanguage{english}
\( \tau _{1} \)
\selectlanguage{british}\\
\cline{2-2} \cline{3-3} \cline{4-4} \cline{5-5} 
\multicolumn{1}{|c|}{1}&
\selectlanguage{english}
0.011
\selectlanguage{british}&
\selectlanguage{english}
4.92
\selectlanguage{british}&
\selectlanguage{english}
1.40
\selectlanguage{british}&
\selectlanguage{english}
5.69
\selectlanguage{british}\\
\hline 
\multicolumn{1}{|c|}{component}&
\selectlanguage{english}
\( p_{2} \)
\selectlanguage{british}&
\selectlanguage{english}
\( t_{2} \)
\selectlanguage{british}&
\selectlanguage{english}
\( \sigma _{2} \)
\selectlanguage{british}&
\selectlanguage{english}
\( \tau _{2} \)
\selectlanguage{british}\\
\cline{2-2} \cline{3-3} \cline{4-4} \cline{5-5} 
\multicolumn{1}{|c|}{2}&
\selectlanguage{english}
0.027
\selectlanguage{british}&
\selectlanguage{english}
10.1
\selectlanguage{british}&
\selectlanguage{english}
2.73
\selectlanguage{british}&
\selectlanguage{english}
16.8
\selectlanguage{british}\\
\hline 
\multicolumn{1}{|c|}{elastic}&
\selectlanguage{english}
\( p_{el} \)
\selectlanguage{british}&
\selectlanguage{english}
\( t_{el} \)
\selectlanguage{british}&
\selectlanguage{english}
\( \sigma _{el} \)
\selectlanguage{british}&
\selectlanguage{english}
\( \tau _{el} \)
\selectlanguage{british}\\
\cline{2-2} \cline{3-3} \cline{4-4} \cline{5-5} 
\multicolumn{1}{|c|}{component}&
\selectlanguage{english}
0.024
\selectlanguage{british}&
\selectlanguage{english}
51.0
\selectlanguage{british}&
\selectlanguage{english}
2.89
\selectlanguage{british}&
\selectlanguage{english}
-6.52
\selectlanguage{british}\\
\hline
\end{tabular}\par}
\end{table*}

The peak with probability \( p_{r}=5.8\cdot 10^{-4} \) at \( t=20.23 \)
ns is caused by the light feedback on the laser optical splitter system.
The light guide of about 1.5 m delivers photons from the laser head
to the optical splitter. On the output of the light guide there is
a lens focusing light on the bunch of light guides, which in turn
are delivering light to the dark room. In order to provide a single
photoelectron regime, a reflective attenuation filter is placed between
the lens and fibers input. The filter reflects part of the light,
which after traveling back and forth between the filter and the laser
head can be fed back to the system%
\footnote{The light splitting system has been manufactured by independent professional
and has been used as is. The reflection peak can be seen only with
a high statistics data, and was noted only aftet the final data processing.
The hardware problem could be easily solved rotating filter by a small
angle.
}. The spread of the peak \( \sigma =1.07 \) ns coincides with a main
peak spread. This peak was extracted from the final shape.

The remaining late pulses shape is described by three functions given
by equation (\ref{Formula:ExpGauss}) type, two with negative and
one with a positive slope (signs in eq. (\ref{Formula:ExpGauss})
are inverted). The position of the last peak helps in clarifying its
origin. The difference between the position of the last peak and the
main peak is \( \Delta t=47.6 \) ns%
\footnote{We are using here results of the separate fit of the position of the
last peak with a gaussian. The parameter \( t_{el} \) from the Table
\ref{Table:LatePulsesFit} can't be used in this estimation, because
the model function (\ref{Formula:ExpGauss}) gives .. of the many
individual contributions. For example the function of the form (\ref{Formula:ExpGauss})
can be used to fit the early pulses shape instead of (\ref{Formula:EarlyPulsesSum})
with a same result.
}, and it is in perfect coincidence with a double drift time obtained
in the previous subsection: \( 2t_{d}=47.2 \) ns. The double drift
time can be explained by electrons wich elastically scatter on the
first dynode electrons, then go away from the dynode, slows down and
stops near the photocathode, and then go back to the first dynode
to produce a signal. The amplitude of this pulses should be the same
as that of the main peak pulses, which is confirmed in \cite{PreLatePulses}
by measuring the transit time of the PMT with a higher threshold.
The total probability to observe elastically scattered photoelectron
is \( p_{el}=0.024 \).

Two remaining contributes with a negative slope corresponds to an
inelastic scattering of the photoelectron on the first dynode, without
any secondaries produced. In this case, part of initial energy of
the incident electron is dissipated as heat in the material of the
dynode, and the drift time of the electron in this case depends on
the remaining part of the energy, and, naturally, is less than in
the case of elastic scattering. In the extreme case all the energy
is dissipated, and, without any delay, the electron is transferred
to the next stage of amplification, producing on average a signal
with an amplitude of factor \( g_{1} \) smaller than a normal signal.
In the intermediate case, the scattered electron is delayed by the
time in the range \( 0-2t_{d} \), and after returning back to the
first dynode produces a signal with lower amplitude in comparison
to the amplitudes of the main peak signals. The smaller is the delay
the smaller is the amplitude of the signal.

The total amount of the inelastically scattered photoelectrons can
be defined summing these two contributes to the late pulses. The summing
gives a value \( p_{in}=0.038 \), i.e. almost 4\% of all registered
signals are the signals of small amplitudes due to inelastically scattered
photoelectrons. This value is less than a value of 6.4\% obtained
in section \ref{Section:Testing} using the values of \( p_{U} \)
, \( A \) and the TDC threshold from Table \ref{Table:Tests}. A
contribution of 8.4\% from the main peak fit..... This observations
leads to a conclusion that the underamplified part of the signals
is mainly due to the totally inelastic scattering on the first dynode
with a minimal delay. The amplitude of the pulses arriving at \( t>5 \)
ns is bigger than a threshold set. In fact, when fitting the charge
spectrum of the PMT with a sum of exponential and gaussian terms,
the valley between them remains underfilled.

\begin{figure*}[!thhhh]
{\centering \resizebox*{1\textwidth}{0.5\textheight}{\includegraphics{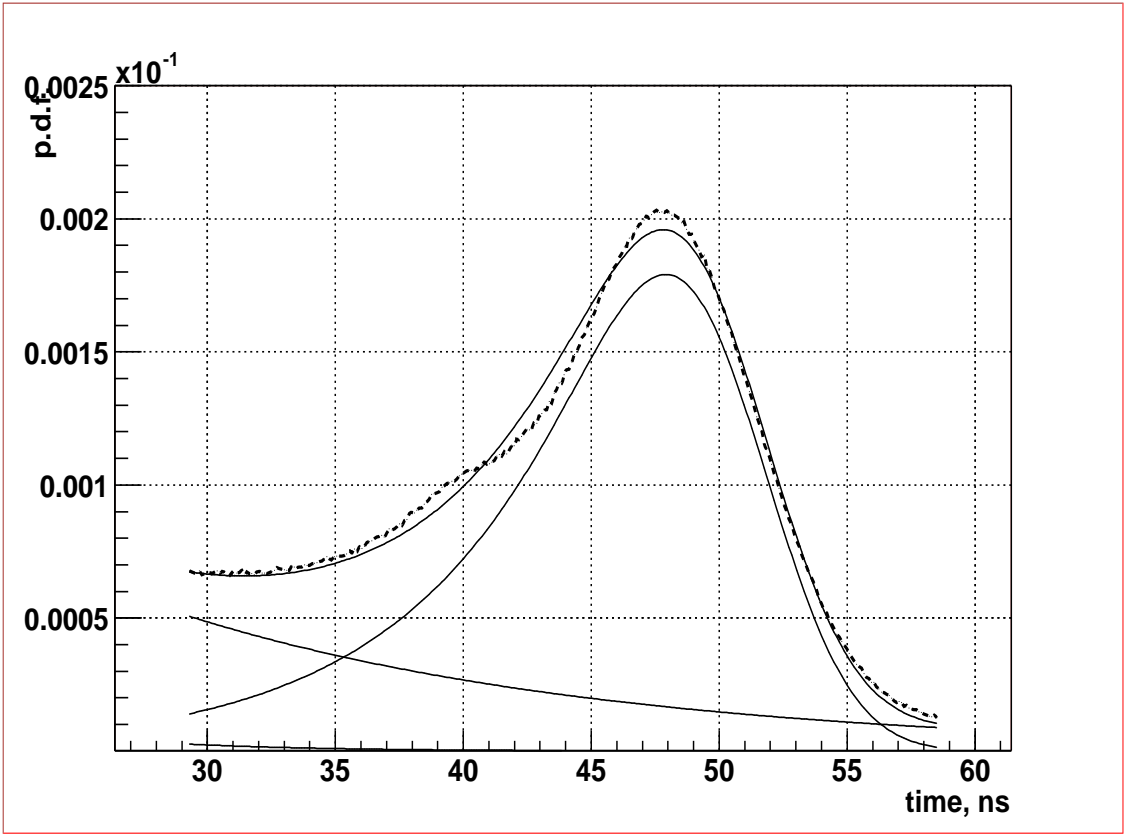}} \par}

\caption{\label{Figure:Bump}The bump on the transit time curve can be clearly
seen at 41 ns.}
\end{figure*}

The proposed model fits well the observed distribution; nevertheless,
the statistics are so high that some further features of the transit
time of elastically scattered photoelectrons can be noted. At the
increasing part of the elastic peak there is a small bump in the region
of 41 ns (see Fig.\ref{Figure:Bump}). The amount of this pulses is
very small, but the bump is pronounced. These are the photoelectrons,
elastically scattered from the focusing grid before the first dynode.
These electrons can reach the photocathode, and after the elastic
scattering on the photocathode, or possible multiplication on it it
will arrive to the first dynode with an energy of the normally accelerated
electrons. In the case of multiplication on the photocathode this
results in a bigger amplitude signals. This can partially explain
the non-gaussian tails observed in the charge distribution of the
single photoelectron signals.

\section{Discussions}

The method proposed for the deconvolution of the PMT signal can be
used as well to extract the time decay curve of the liquid scintillator,
using the data obtained with the start-stop measurements with TDC.

No signals have been observed at the position corresponding to the
single drift time of photoelectron \( t_{d} \), that would have been
present in the case of generation of luminiscent photons or gamma-rays
on the first dynode by an incident photoelectron, without producing
secondaries. The presense of the small peak, nearly at the same position
due to the reflections in the light splitting system gives a possibility
to estimate the sensitivity of our setup to this kind of process at
the level of \( 10^{-5} \) (one should note, that the peak with a
probability \( 5.8\cdot 10^{-4} \) is clearly seen, and the position
of the hypothetical peak is known). 

\begin{figure*}[!thhhh]
{\centering \resizebox*{1\textwidth}{0.5\textheight}{\includegraphics{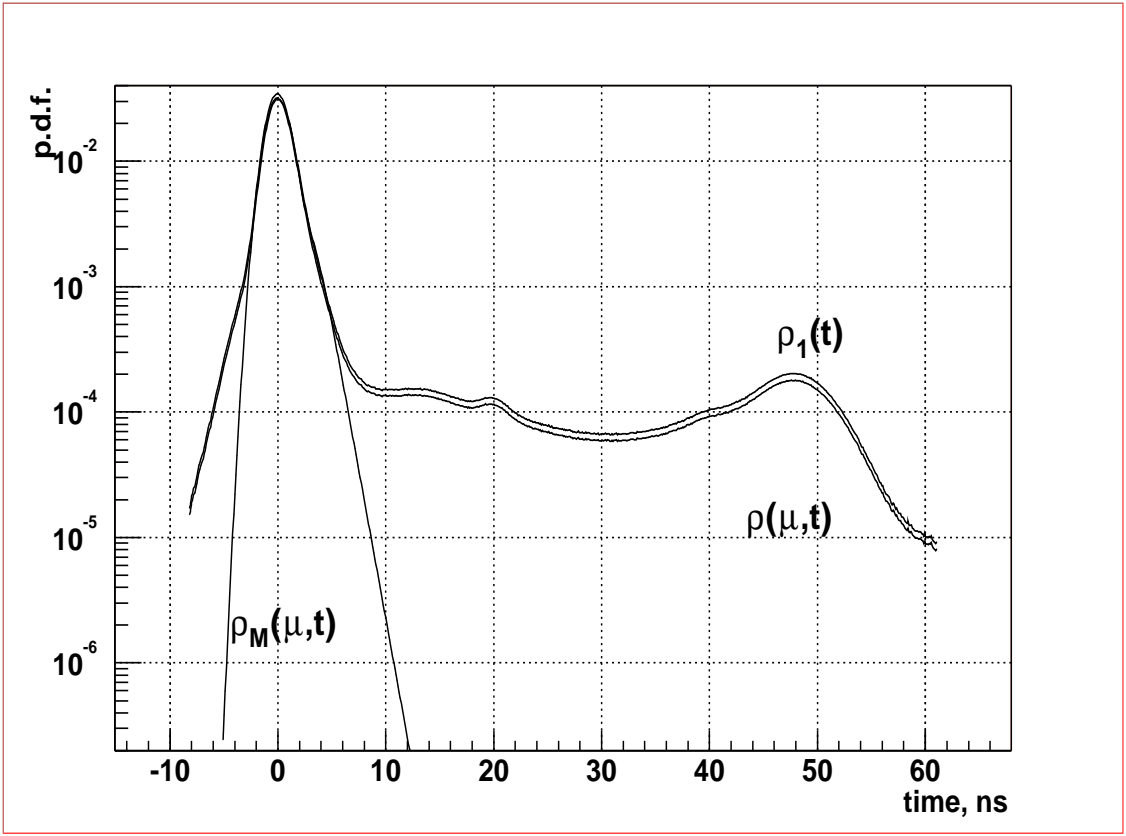}} \par}

\caption{\label{Figure:LowMu}The modeled PMT response to the \protect\( \mu =0.25\protect \)
p.e. light source \protect\( \rho (\mu ,t)\protect \), the single
photoelectron response \protect\( \rho _{1}(t)\protect \) and the
model of the PMT response with only two terms of the main peak function
(without early pulses), \protect\( \rho _{M}(\mu ,t)\protect \) .}
\end{figure*}

The good knowledge of the PMT timing response can help the manufacturer
in improving the PMT timing characteristics. From the point of view
of the experimentor, a good knowledge of the PMT response is necessary
for the proper modeling of the detector response, and in most practical
cases one can simplify the model, keeping only the main contributions
to the signal. In Fig.\ref{Figure:LowMu} are shown PMT responses
modeled with a \( \rho _{1}(t) \) function and with a function \( M(t) \)
without the early pulses term. One can see that the function \( M(t) \)
fails to describe the PMT timing response, and the function \( \rho _{1}(t) \)
practically coincide with \( \rho (\mu ,t) \), reflecting the fact,
that the PMT operates mainly in a single p.e. regime.

\begin{figure*}[!thhhh]
{\centering \resizebox*{1\textwidth}{0.5\textheight}{\includegraphics{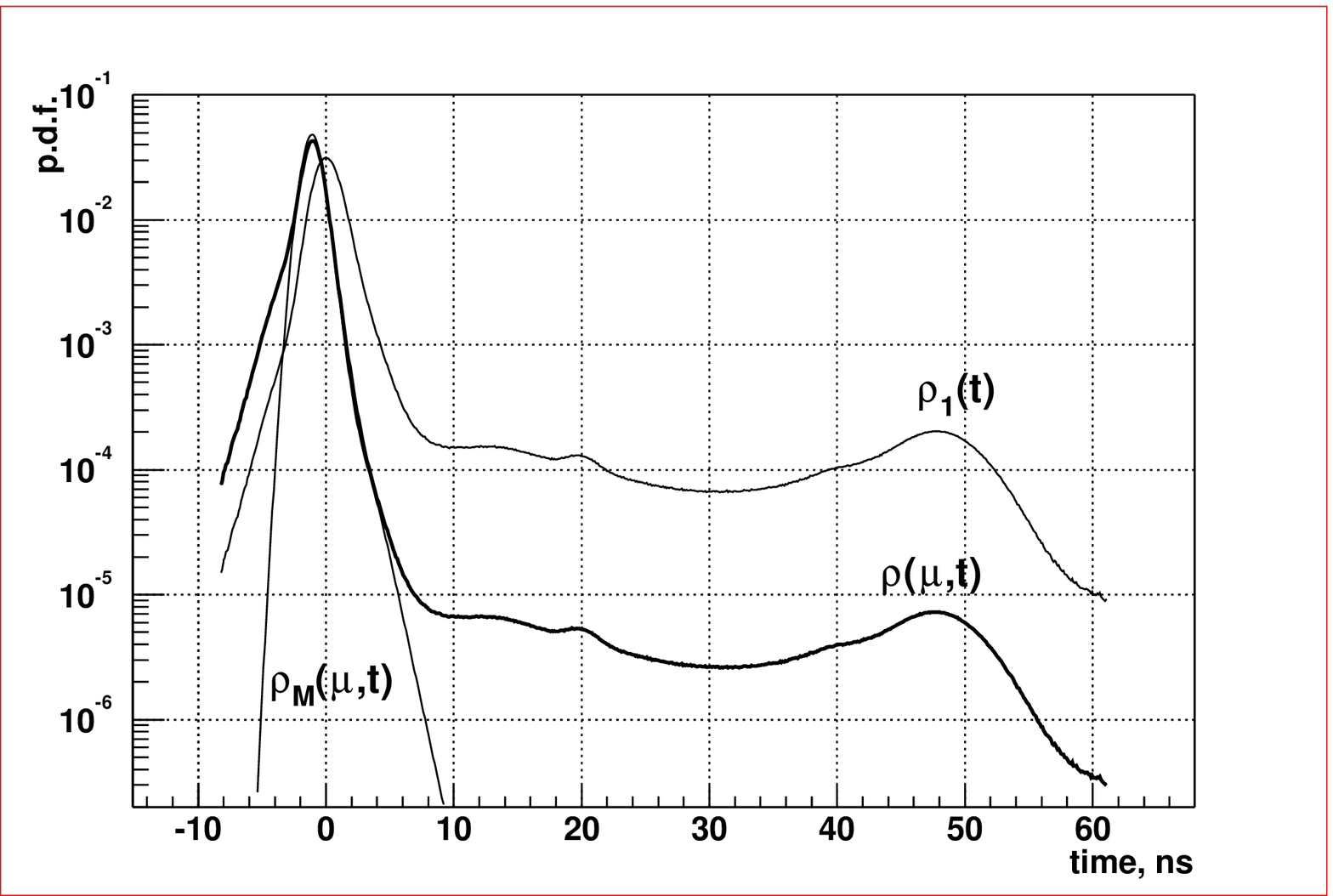}} \par}

\caption{\label{Figure:5pe}The modeled PMT response to the \protect\( \mu =5\protect \)
p.e. light source, \protect\( \rho (\mu ,t)\protect \), the single
photoelectron response \protect\( \rho _{1}(t)\protect \) and the
model of the PMT response with only two terms of the main peak function
(without early pulses), \protect\( \rho _{M}(\mu ,t)\protect \).}
\end{figure*}

Another case is illustrated in Fig.\ref{Figure:5pe}. The PMT is registering
on average \( \mu =5 \) p.e. The tail in the PMT transit time distribution
is suppressed by more than one order of magnitude, and the \( \rho _{M}(t) \)
function gives a satisfactory description of the \( t>0 \) ns part
of the distribution. But this time, the early pulses should be taken
into account in order to have a good model of the early coming pulses.

We can conclude, that the multiple p.e. signal can be modeled with
a main peak part of the total distribution, given by equation (\ref{Formula:MainPeak}).
In the case of a single photoelectron counting the best result can
be obtained using the \( \rho _{1}(t) \) function.

The prepulses at \( t=-24 \) ns stays far away from the main peak,
and with a proper treatment can be easily separated. The amplitude
of these pulses is small, and they can be suppressed increasing the
threshold, as it was demonstrated in \cite{PreLatePulses}. In general,
the relative probability of prepulses increases almost linearly with
an increase of the mean number of the registered photoelectrons: \( p_{pp}\simeq \mu \cdot p_{pp}^{1} \),
where \( p_{pp}^{1} \) is a relative probability of the early pulses
in the single p.e. regime. At \( \mu =10 \) p.e. it is still of the
order of \( 10^{-3} \).

\section{Conclusions}

The results of the test measurements of the characteristics of 2200
PMT for the Borexino experiment provide the most complete information
for the evaluation of the ETL9351 timing characteristics with a high
precision. The unique timing characteristics of the setup used and
a huge statistics accumulated during the tests of the PMTs to be used
in the future Borexino experiment, allow to resolve the fine structure
of the PMT timing response.

A method to obtain the probability density function of the single
photoelectron counting from the experimental data is proposed and
applied to derive the PMT average characteristics. For the first time
an analytical model of the single photoelectron time arrival in a
PMT is proposed, describing all the features of the single photoelectron
time arrival. The origin of the small amplitude pulses, as well as
a non-gaussian tail in the amplitude response of PMT are explained.

\section{Acknowledgements}

Credits are given to the developers of the CERN ROOT program \cite{ROOT},
that was used in the calculations and to create all the figures of
the article. Special thanks to R.Ford for the careful reading of the
manuscript.

\end{document}